\documentclass[aps,amsmath, prb,twocolumn,superscriptaddress]{revtex4-2}
\usepackage{lineno}

\usepackage{graphicx}
\usepackage{dcolumn}
\usepackage{bm}
\usepackage{lipsum}
\usepackage{color}
\usepackage{amsfonts,eucal,mathrsfs,amssymb,gensymb}
\usepackage{comment}
\usepackage[table]{xcolor}
\graphicspath{{./figs/}}
\usepackage{multirow}
\usepackage[breaklinks=true,unicode=true,urlcolor = blue,colorlinks = true,citecolor = blue,linkcolor = blue]{hyperref}
\DeclareMathOperator{\sech}{sech}

\renewcommand{\vec}[1]{\bm{#1}}

\def \Kravchuk #1{\textcolor{orange}{#1}} 

\begin{document}

\preprint{}

\title{Propagation and localization of spin excitations at altermagnetic domain walls}

\author{Oksana Peschanska}
\affiliation{Bogolyubov Institute for Theoretical Physics of the National Academy of Sciences of Ukraine, 03143 Kyiv, Ukraine}


\author{Jeroen van den Brink}
\affiliation{Institute for Theoretical Solid State Physics, Leibniz Institute for Solid State and Materials Research Dresden, D-01069 Dresden, Germany}
\affiliation{W{\"u}rzburg-Dresden  Cluster of Excellence, 01062 Dresden, Germany} 
\affiliation{Institute of Theoretical Physics, Technische Universit{\"a}t Dresden, 01062 Dresden, Germany}

\author{Volodymyr P. Kravchuk}
\email{v.kravchuk@ifw-dresden.de}
\affiliation{Bogolyubov Institute for Theoretical Physics of the National Academy of Sciences of Ukraine, 03143 Kyiv, Ukraine}
\affiliation{Institute for Theoretical Solid State Physics, Leibniz Institute for Solid State and Materials Research Dresden, D-01069 Dresden, Germany}
\affiliation{W{\"u}rzburg-Dresden  Cluster of Excellence, 01062 Dresden, Germany}

\begin{abstract}
Altermagnets (A$\ell$Ms) are spin-compensated materials in which opposite-spin sublattices are connected by a symmetry that causes a spin splitting in their elementary excitations.
As there is a strong effect of altermagnetism on domain wall properties, it is quite natural to also expect an enrichment of the physics of magnetic excitations at A$\ell$M domain walls. 
Here, we consider the propagation of spin eigen-excitations along domain walls in easy-axial $d$-wave A$\ell$Ms. Investigating the presence of bound states localized on a domain wall, we find that the effect of the A$\ell$M on the bound states strongly depends on the orientation of the domain wall relative to the crystallographic directions. If the domain wall is oriented along a nodal direction [100] or [010], A$\ell$M does not change the number of bound states; however, it leads to a nonlinear dispersion and a tilt of the wavefront. The effect of A$\ell$M is strongest when the domain wall is oriented along the directions [110] or [$\bar{1}$10], i.e., along the directions of the strongest A$\ell$M splitting in the magnon spectrum. In this case, (i) the additional gapped bound states appear, (ii) degeneracy of the eigenstates with respect to their polarization (right-handed or left-handed precession of the N{\'e}el vector) is removed, and (iii) the localization area of the bound states strongly depends on the eigenfrequency. The latter may lead to strong localization of the bound state at the domain wall. We further consider the influence of a static magnetic field that is applied along the easy axis, and find that the magnetic field induces an asymmetry between the localization regions on opposite sides of the domain wall and sets an upper limit on the absolute value of the propagating eigenstate's wave vector. 
\end{abstract}

\maketitle

\section{Introduction}
Domain walls (DWs) are well-studied~\cite{Hubert98,Kosevich90} one-dimensional topological solitons, which appear as boundaries between different domains in systems with degenerate ground states. E.g., in easy-axial ferro- and antiferromagnets (AFM), 180$\degree$ DWs separate domains with opposite orientations of the order parameter, which is the magnetization and N{\'e}el vector, respectively. DWs exhibit rich internal dynamics that, in the linear regime, can be related to bound spin eigenstates. The latter are well studied for both ferro-\cite{Winter61,Thiele73a,Braun94,Zhang18c} and AFM~\cite{Paul62,Lemmens86,Ivanov95e,Bogdan14,Kim14a,Shen20a,Park21} DWs. In both cases, the DWs are widely considered as natural waveguides in spintronic devices~\cite{Garcia-Sanchez15,Xing16,Lan15,Henry19,Qiu22,Wang25}. In particular, the DW-based spin-wave diode and spin-wave logic hardware architecture was proposed~\cite{Lan15}. In addition to the numerous theoretical studies, it is worth noting a number of successful experimental realizations of the DW-based spin-wave waveguides~\cite{Wagner16,Albisetti18a,Sluka19}.

In the context of energy-efficient magnon-based spintronic devices~\cite{Chumak15}, antiferromagnetically ordered systems have several advantages over ferromagnetic ones, e.g., the high-speed dynamics in the THz range, and the absence of stray fields. It is known that the AFM DW possesses a single bound state, which propagates with a gapless linear dispersion along the wall, and a group velocity that may be very high $\sim10^4$~m/s~\cite{Park21}. The gaplessness reflects the translational invariance of the DW position, i.e., the bound state transforms into a zero-energy translational (Goldstone) mode if the wave vector along the DW becomes very small and ultimately vanishes.

As compensated magnets, altermagnets~\cite{Smejkal22a,Smejkal22b} have the same technological advantages as antiferromagnets. However, due to the more complicated symmetry transformation connecting two sublattices~\cite{Smejkal20,Smejkal22}, they demonstrate richer physics. It has been shown recently that a static A$\ell$M DW acquires a magnetization that strongly depends on the orientation of the DW relative to crystallographic axes~\cite{Gomonay24a,Kravchuk25a}. Since altermagnetism breaks the Lorenz invariance of the Lagrangian formulated for the N{\'e}el vector~\cite{Gomonay24a,Yershov25,Vakili25}, the deformation of the moving DW is more complicated than Lorentz shrinking~\cite{Gomonay24a}. As a result, the effect of the Walker limit takes place when the altermagnetic DW reaches some critical velocity~\cite{Gomonay24a}. Altermagnetic DWs activate the anomalous Hall effect~\cite{Sorn25} even if in the absence of the DW, it is forbidden by magnetic symmetry. 

Taking into account the strong influence of altermagnetism on the DW properties, it is quite natural to expect an enrichment of the physics of the localized eigenstates of an A$\ell$M DW. In this paper, we provide a systematic theoretical analysis of the bound eigenstates propagating along a DW in an easy-axial $d$-wave altermagnet. Within the paper, we use capital letters $T$, $\vec{R}=X\vec{e}_x+Y\vec{e}_y$ and small letters $t$, $\vec{r}=x\vec{e}_x+y\vec{e}_y$ to denote the dimensional and dimensionless time-space coordinates, respectively.

\section{Model and basic solutions}\label{sec:model}
\begin{figure}
    \centering
    \includegraphics[width=\columnwidth]{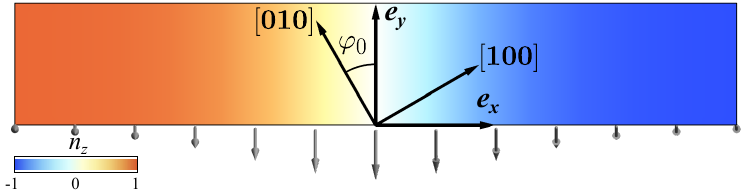}
    \caption{Profile of a static Bloch DW ($\phi_{\text{dw}} = - \pi /2$). Arrows represent N{\'e}el vector orientation, and the density plot shows the distribution of its $z$-component. The reference frame is chosen such that $\vec{e}_x$ and $\vec{e}_y$ are always normal and tangential to the DW, respectively. Orientation of the DW relative to the crystallographic directions is controlled by the angle $\varphi_0$.}\label{fig:DW}
\end{figure}

We base our analysis on the continuum description of A$\ell$Ms, in which the magnetization of each sublattice is represented by a continuous vector field $\vec{M}_\nu(T,\vec{R})$ of constant amplitude $|\vec{M}_\nu|=M_\text{s}$. Here $\nu=1,2$. In the limit $|\vec{M}_1+\vec{M}_2|\ll|\vec{M}_1-\vec{M}_2|$ (so-called exchange approximation), the magnetization $\vec{m}=(\vec{M}_1+\vec{M}_2)/(2M_{\text{s}})\ll1$ is excluded from the equations of motion, and the dynamics of the magnetic subsystem can be presented in terms of the unit N{\'e}el vector field $\vec{n}(T,\vec{R})=(\vec{M}_1-\vec{M}_2)/(2M_{\text{s}})$~\cite{Baryakhtar79,Andreev80,Tveten13}. For the case of $d$-wave A$\ell$Ms, dynamics of $\vec{n}(T,\vec{R})$ is described by Lagrangian ~\cite{Gomonay24a,Yershov25,Vakili25}
\begin{align}\label{eq:L}
		\nonumber\mathcal{L}=&\frac{M_\text{s}}{\gamma^2B_{\text{ex}}}\left[\dot{\vec{n}}+\vec{n}\times\gamma\vec{B}\right]^2\!-\!A_{\text{afm}}\,\partial_\alpha\vec{n}\cdot\partial_\alpha\vec{n}\!-\!K[\vec{n}\times\vec{e}_{z}]^2\\
        &+\frac{A_{\text{alt}}}{\gamma B_{\text{ex}}}\left[(\dot{\vec{n}}+\vec{n}\times\gamma\vec{B})\times\vec{n}\right]\cdot\hat{\mathfrak{D}}_{\{X,Y\}}\vec{n}
\end{align}
supplemented with the constraint $|\vec{n}|=1$. It is assumed here that the effective exchange field $B_{\text{ex}}$ keeping two sublattices antiparallel much exceeds the external field $B$ and the other effective fields in the system. In Eq.~\eqref{eq:L}, $\gamma>0$ denotes the gyromagnetic ratio of an electron, $A_{\text{afm}}$ is the antiferromagnetic stiffness, $K>0$ is the coefficient of the easy-axial anisotropy, $A_{\text{alt}}$ determines the strength of the A$\ell$M effects, and the differential operator $\hat{\mathfrak{D}}_{\{X,Y\}}=2\cos2\varphi_0\partial^2_{XY}-\sin2\varphi_0(\partial^2_{XX}-\partial^2_{YY})$ depends on the angle of rotation of the reference frame relative to the crystallographic directions, namely $\vec{e}_{[100]}=\cos\varphi_0\vec{e}_x+\sin\varphi_0\vec{e}_y$, and $\vec{e}_{[010]}=-\sin\varphi_0\vec{e}_x+\cos\varphi_0\vec{e}_y$. For the relations between the phenomenological constants $A_{\text{afm}}$, $A_{\text{alt}}$ and parameters of microscopic models, see Refs.~\cite{Gomonay24a,Yershov25}. 
For a sufficiently small magnetic field, the ground state of Lagrangian~\eqref{eq:L} is the uniform ordering along the easy-axis: $\vec{n}_0=\pm\vec{e}_z$. The double degeneracy of the ground state gives rise to DWs which separate two domains with opposite orientation of $\vec{n}_0$. In A$\ell$Ms, the static and dynamic properties of a DW strongly depend on its orientation relative to the crystallographic directions~\cite{Gomonay24a,Kravchuk25a,Vakili25}. Here, we choose the reference frame such that $\vec{e}_x$ is always perpendicular to the DW, and the arbitrary orientation of the DW relative to the crystallographic directions is controlled by the angle $\varphi_0$, see Fig.~\ref{fig:DW}.

According to \eqref{eq:L}, a static solution is not affected by the altermagnetism if $\vec{B}=\vec{0}$. In this case, the DW has a standard profile $\theta_{\text{dw}}=2\arctan{e^{pX/\ell}}$, $\phi_{\text{dw}}=\text{const}$, where we use the spherical angles $\theta$ and $\phi$ to define the orientation of the unit vector $\vec{n}=\sin\theta\left(\vec{e}_x\cos\phi+\vec{e}_y\sin\phi\right)+ \vec{e}_z \cos\theta$. Here $p=\pm1$ is the DW topological charge, and $\ell=\sqrt{A_{\text{afm}}/K}$ is the DW width. The perpendicular magnetic field $\vec{B}=B\vec{e}_z$ generally changes the profile of an A$\ell$M DW, except in cases when the DW is oriented along the nodal directions, i.e., for $\varphi_0=0,\,\pm\pi/2,\,\pi$. For these particular orientations, the DW profile is the same as for an AFM DW~\cite{Kosevich90}: $\theta_{\text{dw}}=2\arctan{e^{pX/\Delta}}$, where the DW width $\Delta=\ell/\sqrt{1-B^2/B_{\text{sf}}^2}$ increases due to the field. Here $B_{\text{sf}}=\sqrt{B_{\text{ex}}K/M_{\text{s}}}$ is the spin-flop field. In the following, we consider an arbitrary oriented DW if $B=0$, however, we restrict ourselves to the DW orientation along the nodal directions if $B\ne0$. The solutions for the arbitrarily oriented DW in a magnetic field will be considered in our upcoming papers.

Let us now consider the linear excitations of the field $\vec{n}$ on top of the uniform ground state $\vec{n}_0=\mathrm{p}\vec{e}_z$, where $\mathrm{p}=\pm1$. We utilize the classical analog of the Holstein-Primakoff representation in which the deviations of $\vec{n}$ from its equilibrium direction $\vec{n}_0=\sin\theta_0\left(\vec{e}_x\cos\phi_0+\vec{e}_y\sin\phi_0\right)+ \vec{e}_z \cos\theta_0$ are encoded by a complex-valued function $\psi$:
\begin{equation}\label{eq:HP}
    \vec{n}=(1-|\psi|^2)\vec{n}_{0}+\sqrt{2-|\psi|^2}\left(\vec{e}_+\psi+\vec{e}_-\psi^*\right)
\end{equation}
The vectors $\vec{e}_\pm = \frac12(\vec{a} \pm i \vec{b})$ with $\vec{a} = \partial_{\theta_0}\vec{n}_0$ and $\vec{b} = \vec{n}_0 \times \vec{a}$ represent a basis in a plane perpendicular to $\vec{n}_0$. In our case, $\vec{e}_\pm=\frac12(\mathrm{p}\vec{e}_x\pm i\vec{e}_y)$. The complex-valued function $\psi$ introduced in \eqref{eq:HP} encodes deviation of the N{\'e}el vector $\vec{n}$ from its equilibrium direction $\vec{n}_0$. Note that $\psi=0$ for $\vec{n}=\vec{n}_0$~\footnote{Using the properties $\vec{n}_0\cdot\vec{e}_\pm=0$, $\vec{e}_{\pm}\cdot\vec{e}_\pm=0$, and $\vec{e}_{\pm}\cdot\vec{e}_\mp=1/2$, one can derive $\psi=2(\vec{e}_-\cdot\vec{n}_\perp)/\sqrt{1+\vec{n}_0\cdot\vec{n}}$ and $\psi^*=2(\vec{e}_+\cdot\vec{n}_\perp)/\sqrt{1+\vec{n}_0\cdot\vec{n}}$. Here $\vec{n}_\perp=\vec{n}-\vec{n}_0(\vec{n}_0\cdot\vec{n})$ is projection of $\vec{n}$ on the plain perpendicular to $\vec{n}_0$.}.

\begin{figure}
    \centering
    \includegraphics[width=\linewidth]{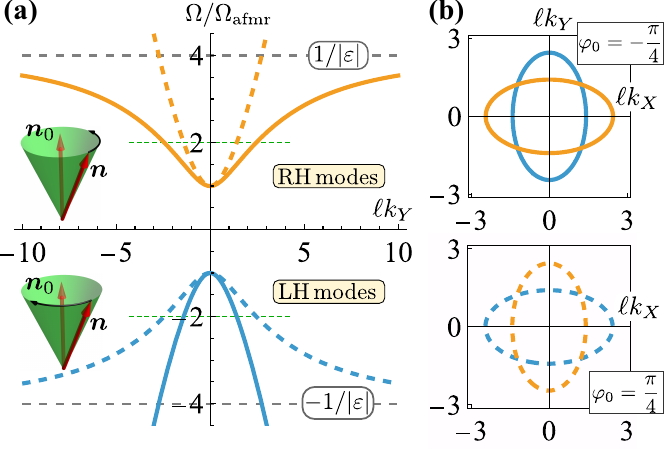}
    \caption{(a) -- Dispersion relations \eqref{eq:disp-sw} for $\varphi_0=-\pi/4$  and $\varphi_0=\pi/4$ are shown by solid and dashed lines, respectively. In both cases, $k_X=0$, and $\varepsilon=0.25$. The positive and negative frequencies correspond to the right-handed and left-handed precession of the N{\'e}el vector above its equilibrium orientation.  (b) -- isosurfaces of constant energies $|\Omega|=2\Omega_{\text{afmr}}$. }\label{fig:disp-$d$-wave}
\end{figure}

The equation of motion linearized with respect to $\psi$ has solution in form of planar waves $\psi \sim e^{i(k_XX + k_YY-\Omega T)}$ with dispersion relation 
\begin{equation}\label{eq:disp-sw}
    \Omega=\left[\pm\sqrt{\Omega^2_{\text{afmr}}+c^2k^2+\Lambda^2k^4\xi_\chi^2}+\Lambda k^2\xi_\chi\right]+\text{p}\gamma B,
\end{equation}
for details see Appendix~\ref{app:GS-DW}. Here $\Omega_{\text{afmr}} = \gamma B_{\text{sf}}$ is frequency of the uniform antiferromagnetic resonance. $c=\ell\Omega_{\text{afmr}}$ is the maximal magnon velocity in the AFM limit, and the parameter $\Lambda=\gamma A_{\text{alt}}/M_{\text{s}}$ represents the strength of the altermagnetic effects. The equivalent expression for the A$\ell$M strength is $\Lambda=\varepsilon\Omega_{\text{afmr}}\ell^2$, where $\varepsilon=(A_{\text{alt}}/A_{\text{afm}})\sqrt{K/(B_{\text{ex}}M_{\text{s}})}$ is the dimensionless form of $\Lambda$ if $\ell$ and $\Omega_{\text{afmr}}^{-1}$ are used for the units of distance and time, respectively. The dimensionless anisotropic parameter $\xi_\chi=\frac12\sin2(\chi-\varphi_0)$ is determined by orientation of the vector $\vec{k}=k(\vec{e}_x\cos\chi+\vec{e}_y\sin\chi)$.  Disperstion \eqref{eq:disp-sw} is analyzed in Fig.~\ref{fig:disp-$d$-wave}. We use different signs of frequency in order to distinguish the polarizations of the magnon modes, namely $\Omega>0$ and $\Omega<0$ stay for the right-handed (RH) and left-handed (LH) precession direction of the vector $\vec{n}$ above its equilibrium direction $\vec{n}_0$, see insets in Fig.~\ref{fig:disp-$d$-wave}(a). The structure of the energy iso-surfaces shown in Fig.~\ref{fig:disp-$d$-wave}(b) clearly demonstrates the $d$-wave character of the model~\eqref{eq:L}. 

The continuous model~\eqref{eq:L} is valid in the long-wave approximation, when the characteristic length scales for the inhomogeneities of $\vec{n}(\vec{r})$ much exceed the interatomic distance $a_0$, i.e., $\ell\gg a_0$ and $ka_0\ll1$. It should also be noted that, since higher-order (fourth-order) gradient terms are neglected in the Lagrangian \eqref{eq:L}, the results obtained are consistent with the initial assumptions provided that contributions of order $k^4$ can be neglected. The latter means $\varepsilon(k\ell)^2\ll1$.

\section{Bound eigenstates of the domain wall in zero magnetic field}\label{sec:non-magnetic}
In the previous section, we introduced the typical scales for time $\Omega^{-1}_{\text{afmr}}$ and length $\ell$. In the following part of this paper, we will use dimensionless time-space coordinates $t=\Omega_{\text{afmr}}T$, $\vec{r}=\vec{R}/\ell$.
Now, Lagrangian \eqref{eq:L} obtains form
\begin{equation}\label{eq:L-new}
    \begin{split}
    \mathscr{L}=&\frac{1}{2}\bigl[\dot{\vec{n}}^2-\partial_\alpha\vec{n}\cdot\partial_\alpha\vec{n}-(1-n_z^2)  -\varepsilon\left[\vec{n}\times\dot{\vec{n}}\right]\hat{\mathfrak{D}}_{\{x,y\}}\vec{n}\bigr],
\end{split}
\end{equation}
where $ \alpha \in \{x, y\}$, and the dimensionless A$\ell$M strength $\varepsilon$ 
is the only parameter which controls the system.  

In order to describe linear excitations on the top of the static DW, we utilize Holstein-Primakoff representation~\eqref{eq:HP}, where $\vec{n}_0(x)$ represents the DW profile, i.e., $\theta_0=\theta_{\mathrm{dw}}=2\arctan{e^{px}}$, where $p=\pm1$ is the DW topological charge~\footnote{In the explicit form $\vec{n}_0(x)=\sech x(\vec{e}_x\cos\phi_0+\vec{e}_y\sin\phi_0)-p\tanh{x}\vec{e}_z$, and $\vec{e}_\pm=-\frac12[(p\tanh{x}\cos\phi_0\pm i\sin\phi_0)\vec{e}_x+(p\tanh{x}\sin\phi_0\mp i\cos\phi_0)\vec{e}_y+\sech x\vec{e}_z]$.}. The linear dynamics is determined by the Lagrangian
\begin{align}\label{eq:L-DW-0}
          &\mathscr{L}_{\text{dw}}=\mathscr{L}_{\text{sw}}-\mathscr{U}_{\text{dw}},\\\nonumber
              &\mathscr{L}_{\text{sw}}= \frac12\left[|\dot\psi|^2-|\vec{\nabla}\psi|^2-|\psi|^2+i\varepsilon\dot{\psi}\hat{\mathfrak{D}}_{\{x,y\}}\psi^*+\text{c.c.}\right],\\\nonumber
              &\mathscr{U}_{\text{dw}}=-\frac{2}{\cosh^2x}\left[|\psi|^2+\frac{i\varepsilon}{8}\sin2\varphi_0(\psi\dot{\psi}^*-\psi^*\dot{\psi})\right],
\end{align}
which one obtains by substituting \eqref{eq:HP} into \eqref{eq:L-new} and keeping the terms no higher than quadratic in $\psi$. Here, the spin-wave part $\mathscr{L}_{\text{sw}}$ describes the magnon dynamics above the uniform ground state $\vec{n}_0 = \pm \vec{e_z}$ and is the partial case of Lagrangian \eqref{eq:L-SW-dim} for the vanishing magnetic field. Note that for zero magnetic field, $\mathscr{L}_{\text{sw}}$ does not depend on the direction of the ground-state polarization. $\mathscr{U}_{\textsc{dw}}$ represents the scattering potential created by the domain wall.  At a large distance from the DW, $\mathscr{U}_{\textsc{dw}}$ vanishes, and the magnon dynamics is described solely by the spin wave Lagrangian $\mathscr{L}_{\textsc{sw}}$. In the AFM limit $\varepsilon=0$, the scattering potential $\mathscr{U}_{\textsc{dw}}$ is reduced to the reflectionless P{\"o}schl-Teller potential, and the theory \eqref{eq:L-DW-0} is reduced to the scattering problem previously considered in Refs.~\onlinecite{Tveten14,Kim14a}. Similarly to the AFM case, $\mathscr{U}_{\textsc{dw}}$ is independent of the DW topological charge $p$ and helicity $\phi_0$. However, the A$\ell$M correction contains the time derivatives $\dot{\psi}$, leading to a dependence of the scattering potential on the magnon energy.

The equation of motion, generated by the Lagrangian \eqref{eq:L-DW-0}, has a solution $\psi=\Phi(x)e^{i(k_yy-\omega t)}$ in the form of a wave propagating along the DW with some wave vector $k_y$ and the eigenfrequency $\omega=\omega(k_y)$. Since DW breaks the translational invariance in $x$-direction, the wave amplitude $\Phi$ depends on $x$. It can be presented in form $\Phi(x)=\Psi(x)e^{iqx}$, where parameter $q =-\varepsilon\omega k_y\cos2\varphi_0/ (1-\varepsilon\omega\sin2\varphi_0)$ determines slope of the wave front, see Fig.~\ref{fig:nodal-disp-0}, and function $\Psi(x)$ is determined by the following Schr{\"o}dinger equation 
\begin{equation}\label{eq:Shrod-0}
     \Psi''+\left(-\kappa^2+\frac{u}{\cosh^2x}\right)\Psi=0.
\end{equation}
Here $\kappa^2 = Q^2 - q^2, \ Q^2=(1-\omega^2+k_y^2 +\varepsilon\omega k_y^2\sin2\varphi_0)/(1-\varepsilon\omega\sin2\varphi_0)$, and $u = (2-\varepsilon\omega\sin\varphi_0\cos\varphi_0)/(1-\varepsilon\omega\sin2\varphi_0)$. Note that the magnitude $u$ of the potential depends on the excitation eigenfrequency $\omega$.
 
Schr{\"o}dinger equation \eqref{eq:Shrod-0} can be transformed into a hypergeometric equation with the substitution $z =\frac{1}{2}(1-\tanh{x})$,
see Ref.~\onlinecite{Lamb80}. The  bounded solutions
\begin{equation}\label{eq:Psi-bnd-gen}
    \Psi_n(x)=\mathcal{C}\,\frac{_2F_1(2\kappa+1+n,-n;1+\kappa;z)}{\cosh^{\kappa} x}
\end{equation}
with $\kappa\ge0$ exist under the condition 
\begin{equation}\label{eq:bound-condition}
\kappa+\frac{1}{2}-\sqrt{u+\frac14}=-n,\qquad n=0,1,2,\dots
\end{equation}
For details, see Appendix~\ref{sec:hypergeom-theory}. Here, the nonnegative integer $n$ numerates the localized modes. Due to the integer argument $-n$, the hypergeometric function $_2F_1$ in the numerator in Eq.~\eqref{eq:Psi-bnd-gen} has the form of a $n$th-order-polynomial of $z$~\cite{NIST10}. The typical size of the localization area is $1/\kappa$. The complex-valued constant $\mathcal{C}$ determines the amplitude and the initial phase of the solution. The bound state condition \eqref{eq:bound-condition} implicitly determines the dispersion relation $\omega(k_y)$ for a given $n$.

In the AFM limit ($\varepsilon=0$) parameter $\varphi_0$ drops out of the problem, and one has $u=2$ and $\kappa=(1-\omega^2+k^2_y)^{1/2}$ for all DW orientations.  The condition \eqref{eq:bound-condition} is reduced to $\kappa=1-n$. For $n=0$, one obtains a localized state $\Psi_0=\mathcal{C}/\cosh x$ with $\kappa=1$ and the dispersion $\omega=\pm|k_y|$, see Fig.~\ref{fig:nodal-disp-0}(a). This is the only localized eigen-excitation of the AFM DW; it was previously studied in Ref.~\onlinecite{Park21}.  For $n=1$ one obtains a half-bound state $\Psi_1=\mathcal{C}\tanh{x}$ with the dispersion relation $\omega=\pm(1+k^2_y)^{1/2}$. The latter coincides with the edge of the continuum of the scattering states shown in Fig.~\ref{fig:nodal-disp-0}(a) by a blue line. Using \eqref{eq:HP}, one transforms the excitation $\psi$ into the DW deformation in terms of the N{\'e}el vector. In this way, one finds that the excitation $\psi_0=\Psi_0e^{i(k_yy-\omega t)}$ corresponds to a sinusoidal deformation of the DW, Fig.~\ref{fig:nodal-disp-0}(b). In the limit $k_y\to0$, the deformation period is infinitely large, which corresponds to the uniform DW displacement. The latter does not cost energy, therefore $\omega\to0$. In other words, the considered excitation is a Goldstone mode related to the continuous translation symmetry in the $x$-direction, spontaneously broken by the DW position.

\subsection{Domain wall oriented in a nodal direction}\label{sec:nodal}

Here we consider cases $\varphi_0=0$ and $\varphi_0=\pi/2$, which correspond to the orientation of the DW in directions [010] and [$\bar{1}$00], respectively. In this case the differential operator in~\eqref{eq:L-new} is reduced to $\hat{\mathfrak{D}}_{\{x,y\}}=2\partial^2_{xy}$ and $\hat{\mathfrak{D}}_{\{x,y\}}=-2\partial^2_{xy}$ for $\varphi_0=0$ and $\varphi_0=\pi/2$, respectively. In what follows, we consider the case $\varphi_0=0$, keeping in mind that the rotation of the DW by $\pi/2$ is equivalent to the sign flip of the A$\ell$T parameter. 

\begin{figure}
    \centering
    \includegraphics[width=\linewidth, height = \linewidth]{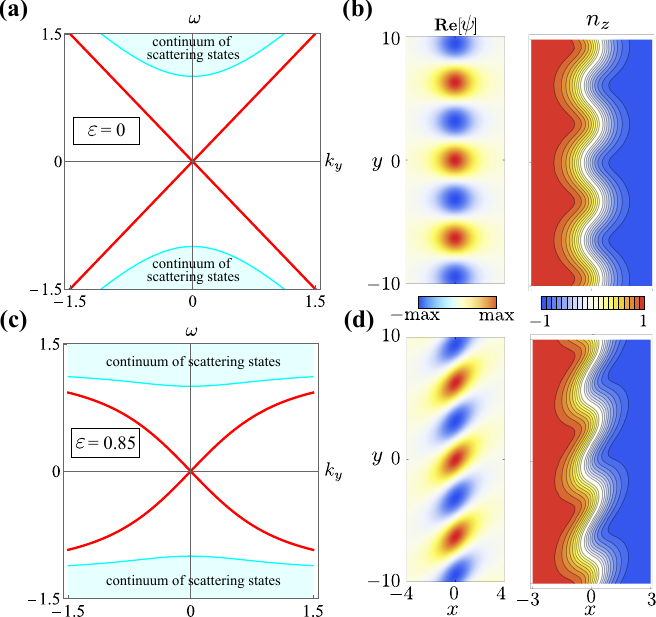}
    \caption{ 
    Dispersion relations for the localized translational mode for the AFM and A$\ell$M DWs oriented in a nodal direction are shown on panels (a) and (c), respectively.
    The distribution of the mode amplitude and the corresponding deformation of the DW profile are shown on panels (b) (AFM), and (d) (A$\ell$T) for $k_y=1$. Here $\varphi_0=0$.
    }\label{fig:nodal-disp-0}
\end{figure}

Similar to the AFM limit, we have $u=2$, and the bound-state condition~\eqref{eq:bound-condition} is reduced to $\kappa=1-n$. Since $\kappa\ge0$, the latter condition can be satisfied only for $n=0$ and $n=1$, which corresponds to the bound ($\kappa=1$), and half-bound ($\kappa=0$) states, respectively. Thus, the corresponding solutions $\Psi_{0}(x)=\mathcal{C}\sech x$ and  $\Psi_{1}(x)=\mathcal{C}\tanh x$ obtained from \eqref{eq:Psi-bnd-gen} are the same as for the AFM limit. In an A$\ell$T, however, the nonzero $q=-\varepsilon\omega k_y$ leads to the slope of the wavefront of the excitation $\psi_{0,1}=\Psi_{0,1}(x)e^{iqx}e^{i(ky-\omega t)}$. The slope angle is $\alpha_{\text{sl}}=\arctan{q/k_y}=-\arctan{\varepsilon\omega}$. This A$\ell$T-induced slope of the wavefront is shown in Fig.~\ref{fig:nodal-disp-0}(d) for a bound solution. The slope of the wavefront is related to the fact that the angle of reflection is not equal to the angle of incidence for planar spin waves that scatter on a DW, which is oriented in a nodal direction~\footnote{Wavefront slope can be intuitively understood within the ray-optic model, in which a bound state propagating along a waveguide is represented as a plane wave that moves in a zigzag pattern between two parallel boundaries~\cite{Saleh91}, where the law of reflection is violated.}; see Section~\ref{sec:scatter}. 

In addition to the wavefront slope, parameter $q$ modifies the dependence $\kappa(\omega,k_y)=[1-\omega^2+k^2_y(1-\varepsilon^2\omega^2)]^{1/2}$, resulting in the modification of the dispersion relations $\omega=\pm|k_y|(1+\varepsilon^2k_y^2)^{-1/2}$ and $\omega=\pm[(1+k^2_y)/(1+\varepsilon^2k_y^2)]^{1/2}$ for the bound and half-bound states, respectively, see Fig.~\ref{fig:nodal-disp-0}(c). As well as in an AFM, the dispersion relation of the half-bound state coincides with the edge of the continuous domain of the scattering states. It is important to note that the considered orientation of DW preserves symmetry between RH- and LH-polarized states.

To summarize, altermagnetism results in two effects on the eigenstates of the DW oriented along the nodal directions: (i) modification of the slope of the wavefront, (ii) nonlinearity of the dispersion relation. The latter is a consequence of the violation of the Lorentz invariance of the Lagrangian~\eqref{eq:L-DW-0} due to the presence of altermagnetic terms.

\subsection{Domain wall oriented in a direction of the maximal altermagnetic splitting}

\begin{figure*}
    \centering
    \includegraphics[width=\linewidth]{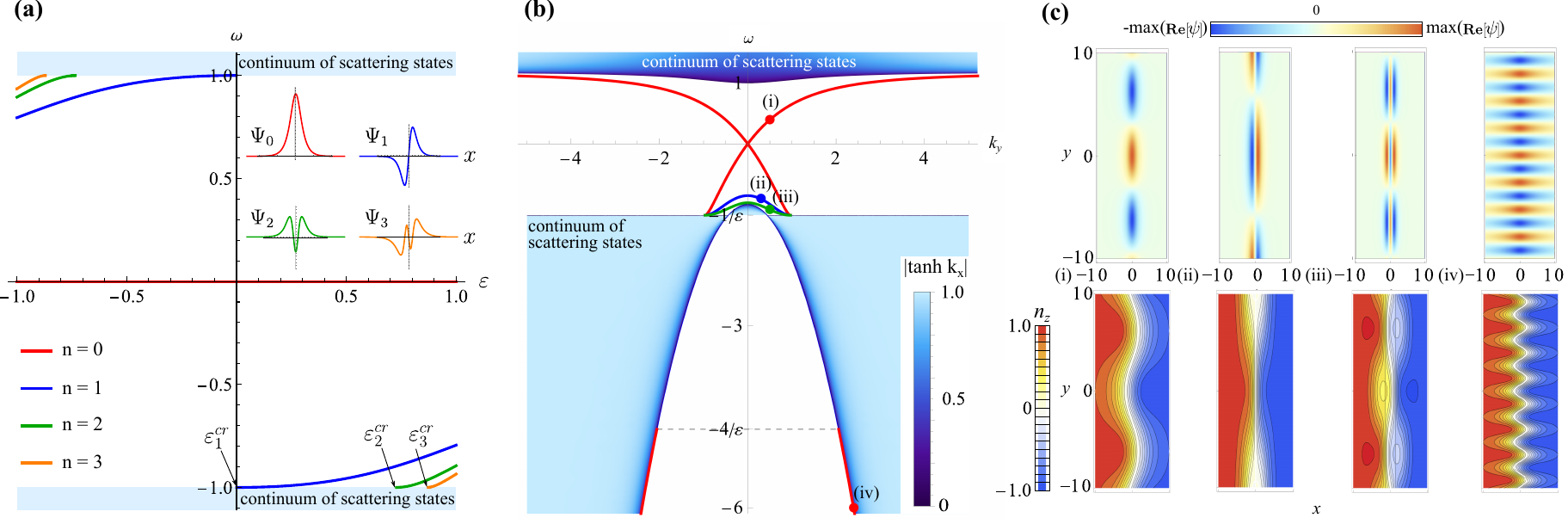}
    \caption{Properties of the bound eigenstates of an A$\ell$T DW oriented in the direction of the maximal A$\ell$T splitting ($\varphi_0=-\pi/4$). Panel (a) shows the dependence of the first four ($n=0,\dots3$) bound-state eigenfrequencies on the value of the A$\ell$T parameter for the case $k_y = 0$; The inset shows the eigenstates' profiles. (b) Dispersion relation of the bound eigenstates for $\varepsilon=0.85$; (c) Examples of the spatial distributions of the eigenstates' amplitude and the corresponding deformation of the DW profile for the points (i) -- (iv) shown on panel (b).}\label{fig:max-disp-0}
\end{figure*}

Here we consider the case $\varphi_0=-\pi/4$, which corresponds to the DW orientation in the direction $[\bar{1}10]$. Properties of the eigenstates of a DW oriented in the perpendicular direction [110] are the same, but with a sign flip of the A$\ell$T parameter $\varepsilon$. For these DW orientations, the directions of the maximal A$\ell$T splitting are parallel and perpendicular to the DW, see Fig.~\ref{fig:disp-$d$-wave}(b). Although the wavefront slope is absent in this case ($q=0$), the altermagnetism significantly enriches the physics of the DW eigenstates, so we start our analysis with the simplest limit $k_y=0$. In this limit, $\kappa=[(1-\omega^2)/(1+\varepsilon\omega)]^{1/2}$ and the bound-state condition \eqref{eq:bound-condition} has the following explicit form
\begin{equation}\label{eq:we-ky0}
    \sqrt{\frac{1-\omega^2}{1+\varepsilon\omega}}=\sqrt{\frac{2+\frac{\varepsilon\omega}{2}}{1+\varepsilon\omega}+\frac14}-n-\frac{1}{2}.
\end{equation}
Relation \eqref{eq:we-ky0} implicitly determines the spectrum $\omega=\omega_n(\varepsilon)$ of the uniform (along DW) bound states, which is shown in Fig.~\ref{fig:max-disp-0}(a). For $n=0$, Eq.~\eqref{eq:we-ky0} has the only solution $\omega\equiv0$, which corresponds to the Goldstone zero mode with profile $\Psi_0(x)=\mathcal{C}\sech x$, see red line in Fig.~\ref{fig:max-disp-0}(a). For $n=1$, solution of Eq.~\eqref{eq:we-ky0} exists for any $\varepsilon$, see blue line in Fig.~\ref{fig:max-disp-0}(a). In the AFM limit $\varepsilon=0$, it coincides with the previously discussed half-bound state with the profile $\Psi_1=\mathcal{C}\tanh x$ and eigenfrequency $|\omega|=1$. In an A$\ell$T with $\varepsilon\ne0$, this state splits off from the edge of the continuum of the scattering states and forms the localized state, whose profile $\Psi_1=\mathcal{C}\tanh x\sech^\kappa x$ has one node at $x=0$. In general case $n\ge1$, Eq.~\eqref{eq:we-ky0} has solution only for $|\varepsilon|\ge\varepsilon^{\text{cr}}_n=(n^2 + n-2)/(n^2 + n-1/2)$. Thus, $\varepsilon_1^{\text{cr}}=0$, $\varepsilon_2^{\text{cr}}=\frac{8}{11}$, $\varepsilon_3^{\text{cr}}=\frac{20}{23}$, ... For the particular case $\varepsilon=\pm\varepsilon^{\text{cr}}_n$, there appear a new half-bound state with $\omega=\mp1$ and $\kappa=0$. For $|\varepsilon|>\varepsilon^{\text{cr}}_n$, this state splits off from the scattering states continuum and localizes in the gap with $|\omega|<1$, $\kappa>0$, and profile $\Psi_n(x)$ containing $n$ nodes, see Fig.~\ref{fig:max-disp-0}(a). For example, $\Psi_2=[(\frac32+\kappa)/(1+\kappa)\tanh^2x-1]\sech^\kappa x$. In the limit $\varepsilon=\pm(\varepsilon^{\text{cr}}_n+\delta\varepsilon)$ with $0<\delta\varepsilon\ll1$, the eigenfrequency  is approximated as $\omega_n\approx\pm(1-a_n\delta\varepsilon^2)$ with $a_n=3/[8(1-\varepsilon^{\text{cr}}_n)^2(3-\varepsilon^{\text{cr}}_n)]$. The corresponding size of the localization area $1/\kappa_n\approx[(1-\varepsilon^{\text{cr}}_n)/(2a_n)]^{1/2}/\delta\varepsilon$ diverges when the localized state approaches the edge of the scattering states continuum. 

For known A$\ell$M materials, $\varepsilon\ll1$. So, the altermagnetically induced bound state with $n=1$ may be the most relevant for practical realizations. For the reason of the smallness of $\varepsilon$, we limit the plot in Fig.~\ref{fig:max-disp-0}(a) to $|\varepsilon|<1$, although the considered eigenstates exist for larger $\varepsilon$ as well.

Let us now consider the generalization with $k_y\ne0$, i.e., the eigenstates propagating along the DW with a finite wavelength. First of all, note the strong modification of the domain of the scattering states, shown by the blue shadowing in Fig.~\ref{fig:max-disp-0}(b). This domain is essentially asymmetric with respect to the RH- and LH-polarized magnons. The domain boundaries can be obtained from the dispersion relation \eqref{eq:disp} for spin waves excited on the top of the uniform state for the case $b=0$ and $\varphi_0=-\pi/4$. Two of these boundaries $\omega_{\text{bnd}}^\pm=\pm[1+k_y^2+\frac14\varepsilon^2k_y^4]^{1/2}-\frac{\varepsilon}{2}k_y^2$ correspond to $k_x=0$, they are shown in Fig.~\ref{fig:disp-$d$-wave}(a) by solid lines. One more boundary $\omega'_{\text{bnd}}=-1/\varepsilon$ corresponds to $k_x\to\pm\infty$.

\begin{figure}[h!]
    \centering
    \includegraphics[width=\linewidth]{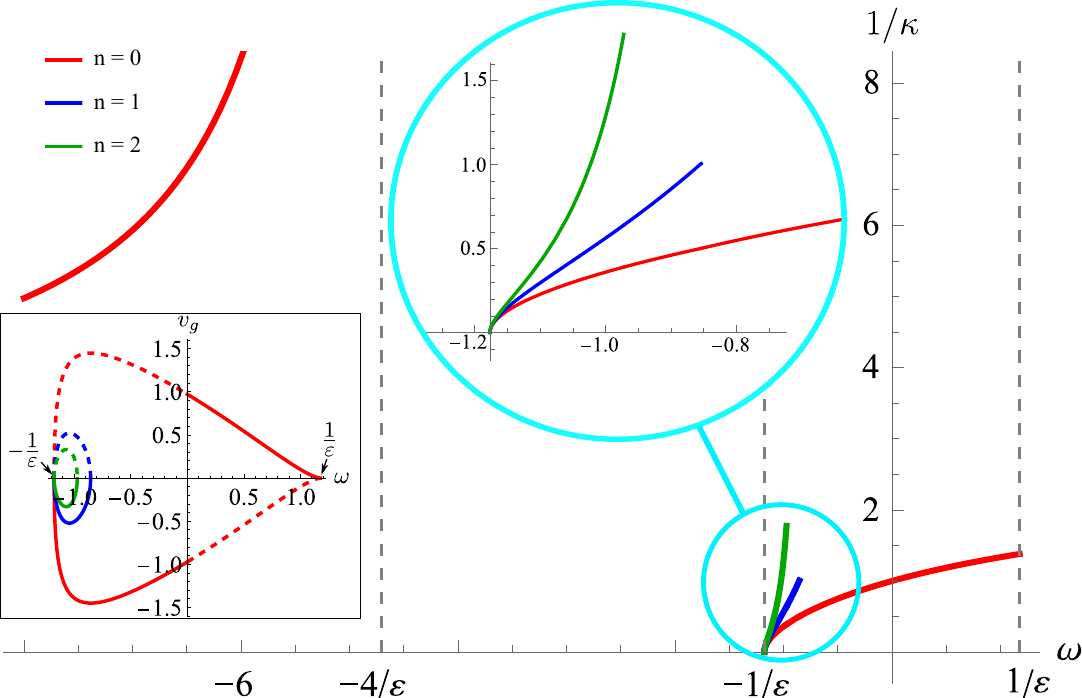}
    \caption{Dependence of the size of the localization region of different bound states on their eigenfrequency for the case $\varphi_0=-\pi/4$, $\varepsilon = 0.85$. The inset shows the frequency-dependence of the group velocity $v_g=\partial_{k_y}\omega$; solid and dashed lines correspond to the cases $k_y>0$ and $k_y<0$, respectively. }\label{fig:loc-reg}
\end{figure}

With the help of \eqref{eq:bound-condition}, we analyze localized modes with a different number of nodes $n$. As previously, for the case $n=0$, we find a translational mode $\psi_0=\sech^\kappa xe^{i(k_yy-\omega t)}$, which demonstrates linear dispersion $\omega\approx\pm|k_y|(1+\varepsilon^2/12)^{-1/2}$ in the long-wave regime ($k_y\ll\varepsilon^{-1}$), see red line in Fig.~\ref{fig:max-disp-0}(b). However, in contrast to the previously considered cases (see Fig.~\ref{fig:nodal-disp-0}), this mode is essentially asymmetric between the RH- and LH-polarizations~\footnote{Note that the properties of the RH- and LH-polarized modes are interchanged if $\varepsilon$ flips the sign.}, and its localization area is not constant, see Fig.~\ref{fig:loc-reg}. While the RH-mode $\psi_0$ exists for arbitrarily large wave vectors, asymptotically approaching the edge $\omega_{\text{bnd}}^+$, its LH counterpart exists only in the interval $|k_y|<k_0=(2+\varepsilon^2)/(4\varepsilon^2)$, touching the boundary $\omega'_{\text{bnd}}$ if $|k_y|=k_0$. Surprisingly, as the mode $\psi_0$ approached the boundary $\omega'_{\text{bnd}}$, it focuses, i.e., its localization size $1/\kappa$ shrinks to zero, see Fig.~\ref{fig:loc-reg}. 
This effect can be explained by the matching of the localization size $1/\kappa$ of the bound state and the wavelength $2\pi/k_x$ of the scattering state at the boundary. Note that $k_x\to\pm\infty$ at $\omega'_{\text{bnd}}$. Modes with $n\ge1$ exist in the same interval $|k_y|<k_0$ of the wave vectors. They also exhibit the focusing effect as they approach the boundary $\omega'_{\text{bnd}}$, see blue and green lines in  Figs.~\ref{fig:max-disp-0}(b) and \ref{fig:loc-reg}. Considering a sharp focusing regime one should keep in mind the restriction $\ell/\kappa\gg a_0$ imposed by the continuous model applicability. Geometrical meaning of the modes with a different number of nodes $n$ is explained in Fig.~\ref{fig:max-disp-0}(c). While the translational mode ($n=0$) corresponds to the sinusoidal bending of the DW, the modes with $n\ge1$ lead to the modulation of the DW width. Note that the high-energy magnon modes responsible for modulating DW width were previously obtained for ring-shaped AFM DWs that model large-radius skyrmions~\cite{Kravchuk19a}.

Remarkably, the LH-polarized translational mode ($n=0$) can exist in the high-energy regime with $\omega<-4/\varepsilon$ for $|k_y|>k'_0=[(16-\varepsilon^2)/(5\varepsilon^2)]^{1/2}$. This mode has a relatively large localization area, see Fig.~\ref{fig:loc-reg}, and it is located in close proximity to the boundary $\omega_{\text{bnd}}^-$ without touching it, see Fig.~\ref{fig:max-disp-0}(b). This mode may be of interest for applications due to its high group velocity $v_g=\partial_{k_y}\omega$. The remaining modes attain a maximum group velocity near the edge frequency $\omega=-1/\varepsilon$ (for $|\varepsilon|\ll1$); see the inset to Fig.~\ref{fig:loc-reg}.






\section{Influence of magnetic field}\label{sec:magnetic}

\begin{figure}
    \centering
    \includegraphics[width=\columnwidth]{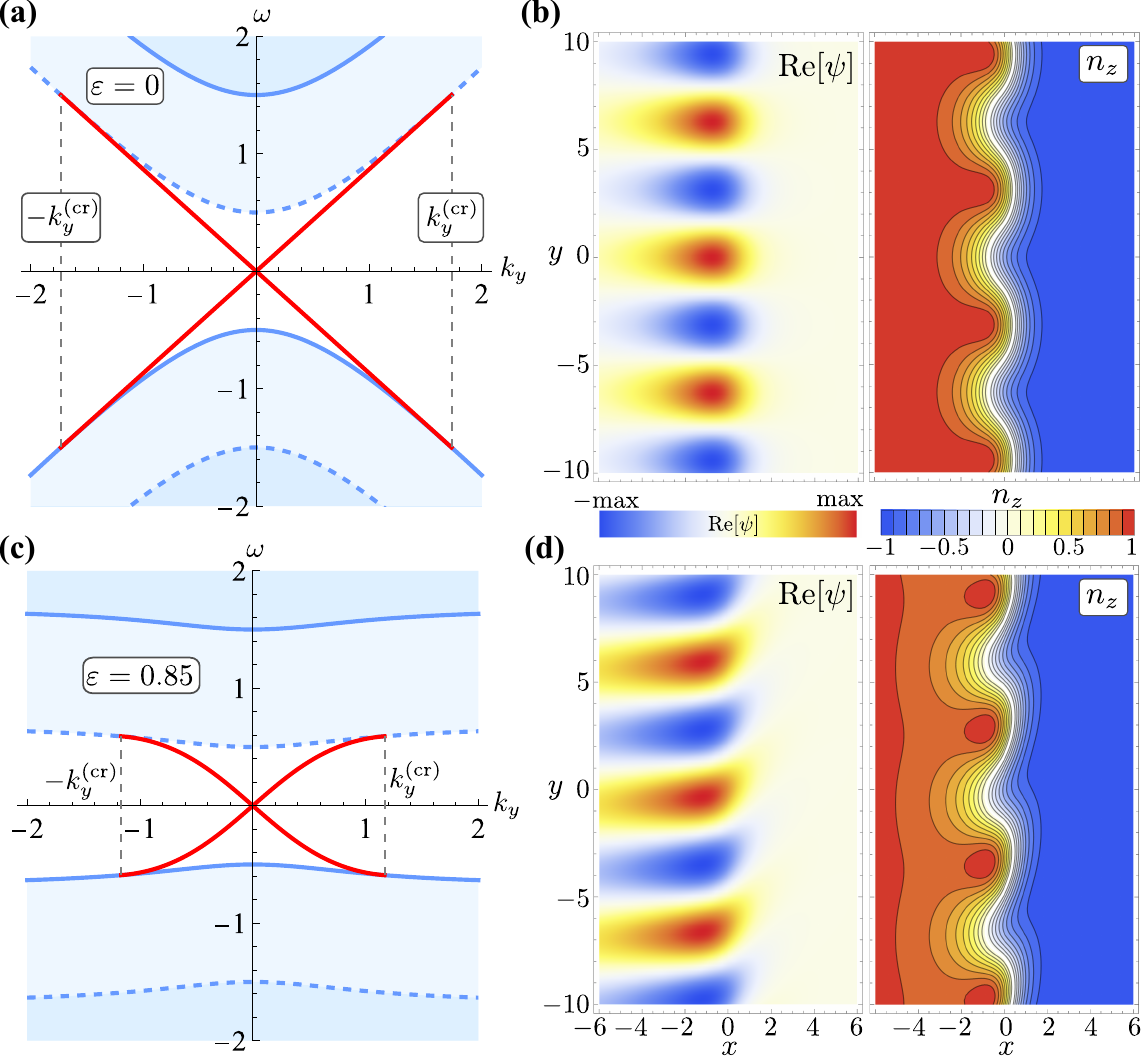}
    \caption{Dispersion relations of the bound state propagating along AFM ($\varepsilon=0$) and A$\ell$T ($\varepsilon=0.85$) DW are shown by red lines on panels (a) and (c), respectively. Solid and dashed blue lines show the edges of the continuum spectrum of magnons over the ground states $\vec{n}_0=\vec{e}_z$ and $\vec{n}_0=-\vec{e}_z$, respectively. Distribution of the magnon amplitude \eqref{eq:Psi-gen} and the corresponding DW deformation are shown on panels (b) and (d) for the AFM ($\varepsilon=0$) and A$\ell$T ($\varepsilon=0.85$) cases, respectively. In both cases, we consider RH-polarized magnons with $k_y=1$. In all cases, $b=0.5$. Here $\varphi_0=0$.}
    \label{fig:ALTM-b}
\end{figure}

In this section, we complete Lagrangian \eqref{eq:L-new} with terms that account for the external magnetic field $\vec{B}=B\vec{e}_z$ oriented along the easy axis~\cite{Yershov25}:

\begin{equation}\label{eq:L-magnetic}
    \begin{split}
    \mathscr{L}=&\frac{1}{2}\{(\dot{\vec{n}}+\vec{n}\times\vec{b})^2-\partial_\alpha\vec{n}\cdot\partial_\alpha\vec{n}-(1-n_z^2) \\
    &-\varepsilon\left[\vec{n}\times(\dot{\vec{n}}+\vec{n}\times \vec{b})\right]\cdot\hat{\mathfrak{D}}_{\{x,y\}}\vec{n}\},
\end{split}
\end{equation}
where $\vec{b}=\vec{B}/B_{\text{sf}}$. Looking for a static one-dimensional solution $\vec{n}_{\text{dw}}=\sin\theta_{\text{dw}}(x)[\vec{e}_x\cos\phi_{\text{dw}}(x)+\vec{e}_y\sin\phi_{\text{dw}}(x)]+\vec{e}_z\cos\theta_{\text{dw}}(x)$ of \eqref{eq:L-magnetic}, we find that the standard DW solution~\cite{Kosevich90} $\theta_{\text{dw}}=2\arctan{e^{p\varkappa x}}$, $\phi_{\text{dw}}=\text{const}$ with $\varkappa=\sqrt{1-b^2}$ and $p=\pm1$ is allowed only for a particular case when the DW is oriented along the nodal directions, e.g., for $\varphi_0=0,\pm\frac{\pi}{2},\pi$. In the general case, both angles $\theta_{\text{dw}}$ and $\phi_{\text{dw}}$ are coordinate dependent. This effect of A$\ell$T-induced deformation of the DW profile in a magnetic field will be studied in our upcoming papers. Here we consider the particular case of the nodal DW orientation.


Using Holstein-Primakoff representation \eqref{eq:HP} with $\vec{n}_0=\vec{n}_{\text{dw}}$, we derive Lagrangian for linear excitations above the DW 
\begin{equation}\label{eq:L-h}
   \begin{split}
   \mathscr{L}_{\text{dw}}^{{\text{b}}}&\!\!=\!\frac{|\dot\psi|^2}{2}-\frac{|\vec{\nabla}\psi|^2}{2}+\frac{1-b^2}{2}\!\left(\frac{2}{\cosh^2\varkappa x}-1\right)\!|\psi|^2\\
   &+i\varepsilon\dot{\psi}\psi^*_{xy}+pb(i\dot{\psi}+\varepsilon\psi_{xy})\psi^*\tanh{\varkappa x}+\text{c.c.}
   \end{split}
\end{equation}
Similar to Section~\ref{sec:nodal}, we consider here the case $\varphi_0=0$, keeping in mind that each rotation of the DW  by $\pi/2$ is equivalent to the sign flip of $\varepsilon$.

We look for a solution $\psi=\Phi(x)e^{i(k_yy-\omega t)}$ in the form of a planar wave traveling along DW. By choosing $\Phi(x)=\Psi(x)e^{iF(x)}$ with $F=qx-pb\varepsilon k_y\varkappa^{-1}\ln\cosh\varkappa x$, we find the amplitude $\Psi(x)$ as a solution of the Schr{\"o}dinger equation with Rosen-Morse potential:

\begin{equation}\label{eq:phi0simp}
    \begin{split}
        & \Psi'' + \varkappa^2\big\{ -\alpha + \beta \sech^2{\varkappa x} +  \gamma \tanh{\varkappa x} \big\} \Psi = 0 ,
    \end{split}
\end{equation}
where $\alpha =  1 - [\omega^2 - k_y^2 (1 - \varepsilon^2  (b^2 + \omega^2))]\varkappa^{-2}, \   \beta = 2-\varepsilon^2 b^2 k_y^2\varkappa^{-2}, \  \gamma = 2 pb\omega(1+\varepsilon^2 k_y^2)\varkappa^{-2}$. In the limit $b=0$, Eq.~\eqref{eq:phi0simp} is reduced to Eq.~\eqref{eq:Shrod-0} for the case $\varphi_0=0$. 
Equation \eqref{eq:phi0simp} appeared in problems of the magnon scattering on precessing~\cite{Bogdan14} and skyrmion-textured~\cite{Lee23} AFM DWs. Similar to Section~\ref{sec:non-magnetic}, we reduce Eq.~\eqref{eq:phi0simp} to the hypergeometric form using the change of variables $z = \frac12(1-\tanh{\varkappa x})$, see Appendix~\ref{sec:hypergeometric-magnetic}. The bounded solutions
\begin{equation}\label{eq:psi-n}
    \Psi_n\!=\!\mathcal{C}z^{\frac{\kappa_+}{2}}(1-z)^{\frac{\kappa_-}{2}}\!{}_2F_1(\kappa_+\!+\kappa_-\!+1+n,-n;1+\kappa_+;z)
\end{equation}
with $\kappa_\pm=\sqrt{\alpha\pm\gamma}$ exist under the condition
\begin{equation}\label{eq:cnd-b}
    \frac{\kappa_+\!+\kappa_-\!+1}{2}-\sqrt{\beta+\frac14}=-n,\qquad n=0,1,\dots,
\end{equation}
for details see Appendix~\ref{sec:hypergeometric-magnetic}. Solution \eqref{eq:psi-n} has asymptotic behavior $\Psi_n\sim e^{\mp\kappa_{\pm}\varkappa x}$ for $x\to\pm\infty$. I.e., the size of the localization region differs on the two sides of the domain wall. For the RH-polarized solutions,  the localization is larger/smaller at the domains where the magnetic field is parallel/antiparallel to the N{\'e}el vector. And it is vice versa for the LH-polarized excitations. For the vanishing magnetic field, one obtains $\kappa_+=\kappa_-=\kappa$, and $\beta=u$, thus the condition \eqref{eq:cnd-b} as well as the solution \eqref{eq:psi-n} are reduced to \eqref{eq:bound-condition} and \eqref{eq:Psi-bnd-gen}, respectively.

The bound states condition \eqref{eq:cnd-b} implicitly determines the dispersion relation $\omega=\omega(k_y)$ for the localized excitations propagating along the DW. In the AFM limit ($\varepsilon=0$), condition \eqref{eq:cnd-b} is reduced to $\frac12(\kappa_++\kappa_-)=1-n$. In a magnetic field ($b\ne0$), the latter can be satisfied only for $n=0$. In this case, the localized state has profile $\Phi=\Psi_0=\mathcal{C}e^{\mp pb|k_y|x}\sech \varkappa x$ and the dispersion relation $\omega=\pm\varkappa|k_y|$. Here, signs `$+$' and `$-$' correspond to RH- and LH-polarizations, respectively. This solution exists for $|k_y|<k_y^{(\text{cr})}=\varkappa/|b|$, see Fig.~\ref{fig:ALTM-b}(a). A DW connects two ground states polarized in directions $\pm\vec{e}_z$. In a magnetic field, the continuum spectra of magnons excited over uniform states $\vec{n}_0=\pm\vec{e}_z$ differ, see Eq.~\eqref{eq:disp}. The dispersion relation of the bound state is sandwiched between the two boundaries of the continuous spectra, corresponding to the domains polarized up and down, see Fig.~\ref{fig:ALTM-b}(a). For $k_y=\pm k^{(\text{cr})}_y$, the dispersion of RH-polarized (LH-polarized) magnons touches the continuum spectrum from the domain polarized antiparallel (parallel) to the field, leading to delocalization on the side of the domain polarized parallel 
(antiparallel) to the field. The asymmetries in the magnon amplitude distribution, as well as in the DW deformation, are shown in Fig.~\ref{fig:ALTM-b}(b) by the example of the RH-polarized bound state. 

In the A$\ell$T case, the behavior of the translational localized mode ($n=0$) is qualitatively the same as in the AFM limit, but the nonlinear dispersion relation, see Fig.~\ref{fig:ALTM-b}(c), and a more complicated eigenfunction
\begin{equation}\label{eq:Psi-gen}
   \psi=\mathcal{C}\frac{e^{-\frac12(\kappa_+-\kappa_-)\varkappa x+iF(x)}}{\cosh^{\frac12(\kappa_++\kappa_-)}\varkappa x }e^{i(k_yy-\omega t)},
\end{equation}
where the dispersion $\omega=\omega(k_y)$ is implicitly determined by Eq.~\eqref{eq:cnd-b} for $n=0$. Due to the additional phase factor $e^{iF(x)}$, the DW deformation possesses additional nonuniformities in the $x$-direction, see Fig.~\ref{fig:ALTM-b}(d). As well as for the case $b=0$, the additional bound states do not appear if the DW is oriented in a nodal direction. In other words, Eq.~\eqref{eq:cnd-b} does not have solutions for $n>0$.

\begin{figure}
    \centering
    \includegraphics[width=\linewidth]{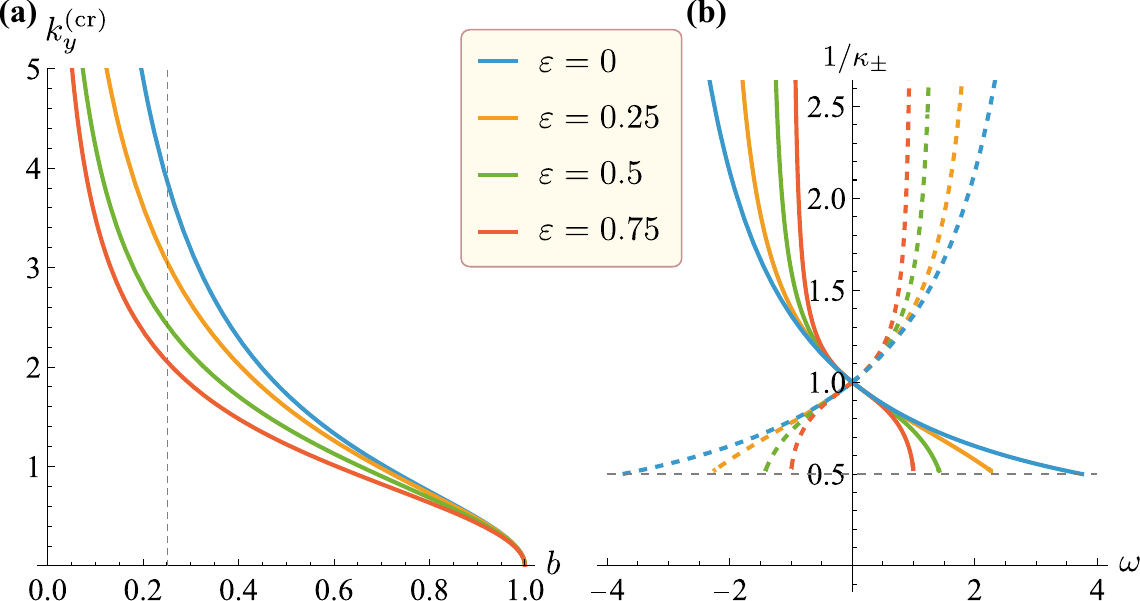}
    \caption{(a) Dependence of the critical value of the wave-vector $k_y^{\text{(cr)}}$ on the magnitude of the magnetic field $b$ for different values of the A$\ell$T parameter $\varepsilon$. (b) Localization regions to the right ($1/\kappa_+$) and to the left ($1/\kappa_-$) of the DW with $p=1$ are shown by the solid and dashed lines, respectively, for the same set of values of $\varepsilon$ and magnetic field $b=0.25$ shown by the dashed vertical line on panel (a).}\label{fig:spct-b}
\end{figure}

The cut-off value $k^{(\text{cr})}_y$ of the wave vector strongly depends on the magnetic field, and altermagnetism makes this dependence steeper, see Fig.~\ref{fig:spct-b}(a). In the limit $|\varepsilon|\ll|b|\ll1$, we estimate $k_y^{(\text{cr})}\approx\varkappa(1-\varepsilon^2/b^2)^{1/2}/|b|$. The influence of altermagnetism on localization regions $1/\kappa_+$ and $1/\kappa_-$ is analogous, see Fig.~\ref{fig:spct-b}(b); i.e., it amplifies the field-induced asymmetry in the spatial distribution of the magnon intensity. It is worth noting that when the delocalization occurs on one side of the DW $1/\kappa_\pm\to\infty$~\footnote{When $k_y\to k_y^{(\text{cr})}$ and $\omega\to\omega^{(\text{cr})}=\omega(k_y^{(\text{cr})})$.}, the localization region on the other side of the DW is $1/\kappa_\mp\gtrapprox1/2$ with $1/\kappa_\mp=1/2$ for $\varepsilon=0$.

\section{Scattering states}\label{sec:scatter}

Let us briefly discuss the main properties of the scattering states for the case $B=0$. The latter are described by Eq.~\eqref{eq:Shrod-0} with $\kappa^2<0$. For this case, the derivation of the scattering states of Eq.~\eqref{eq:Shrod-0} can be found in Ref.~\onlinecite{Lamb80}. Adopting the results for our case, we may write the general solution for $\psi(t,x,y)$ in the form
\begin{equation}\label{eq:psi-sct}
    \psi=\mathcal{C}\,{}_2F_1(a,b;c;z)(\cosh x)^{^{i(k_x-q)}}e^{i(qx+k_yy-\omega t)},
\end{equation}
where $a=\frac12+(u+\frac14)^{1/2}-i(k_x-q)$, $b=\frac12-(u+\frac14)^{1/2}-i(k_x-q)$, $c=1-i(k_x-q)$, see Appendix~\ref{app:sct}, for details. The dispersion relation $\omega=\omega(k_x,k_y)$ of the scattering states coincides with the dispersion relation of the magnons excited on the top of a uniform domain; in zero magnetic field, it is determined by Eqs.~\eqref{eq:disp} and \eqref{eq:disp-sw} for dimensionless and full representations, respectively.

In the limit $x\to\infty$, solution \eqref{eq:psi-sct} has the form of a planar wave, transmitted through the DW $\psi\sim\mathcal{T}e^{i(k_xx+k_yy-\omega t)}$. While in the opposite limit $x\to-\infty$, it is the superposition of the incident and the reflected waves $\psi\sim e^{i(k_xx+k_yy-\omega t)}+\mathcal{R}e^{i(-(k_x-2q)x+k_yy-\omega t)}$. Here, the transmission $\mathcal{T}$ and reflection $\mathcal{R}$ coefficients are as follows
\begin{equation}\label{eq:TR}
    \mathcal{T}=\frac{\Gamma(a)\Gamma(b)}{\Gamma(c)\Gamma(a+b-c)},\;\mathcal{R}=\frac{\Gamma(a)\Gamma(b)\Gamma(c-a-b)}{\Gamma(c-a)\Gamma(c-b)\Gamma(a+b-c)}.
\end{equation}

In the AFM limit, one obtains from \eqref{eq:TR} $\mathcal{T}=(ik_x-1)/(ik_x+1)$ and $\mathcal{R}=0$; and general solution \eqref{eq:psi-sct} is reduced to
$\psi=(ik_x-\tanh x)/(ik_x+1)e^{i(ik_xx+k_yy-\omega t)}$~\footnote{With the proper choice of the coefficient $\mathcal{C}$.}. The obtained eigenstate, as well as the non-reflecting property of a static AFM domain wall, are well known and are widely discussed in the literature~\cite{Yan11,Kim14a,Tveten14}. 

For an A$\ell$M DW, generally $\mathcal{R}\ne0$ and $\mathcal{R}$ is not an analytical function of $\varepsilon$ at the point $\varepsilon=0$. In the limit $|k_x|\gg|\varepsilon|$ and $|\varepsilon|\ll1$, one finds that
\begin{equation}\label{eq:R-approx}
    \mathcal{R}\approx\pm\frac{i\varepsilon\sin2\varphi_0}{\sinh(\pi k_x)}\frac{\pi}{2}\frac{1-ik_x}{1+ik_x}\sqrt{1+k_x^2+k_y^2},
\end{equation}
where signs `$+$' and `$-$' correspond to RH- and LH-polarized magnons, respectively. From \eqref{eq:R-approx}, we conclude that an A$\ell$M DM oriented in one of the nodal directions still creates a reflectionless potential for spin waves. 

Maximum reflection is achieved when the DW is oriented in the direction of the maximum A$\ell$M splitting of magnons. Let’s examine this case in more detail using the example of a DM oriented in the direction [$\bar{1}$10] ($\varphi_0=-\pi/4$). 
\begin{figure}
    \centering
    \includegraphics[width=\columnwidth]{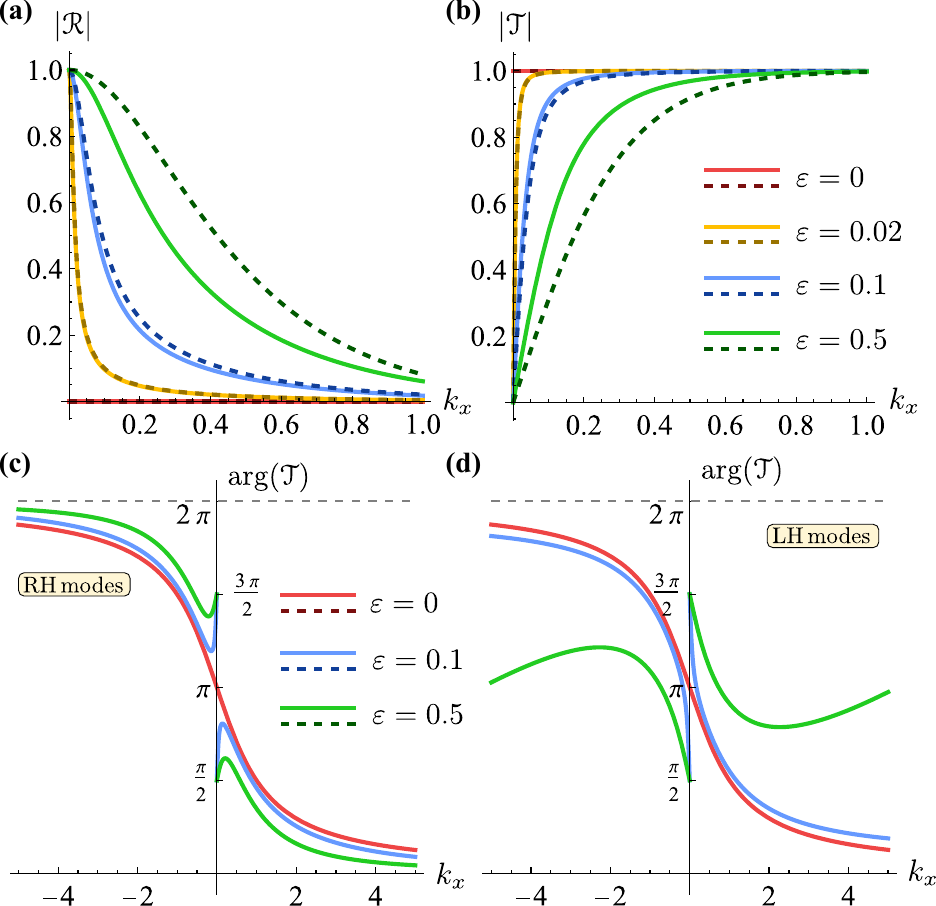}
    \caption{Properties of the reflection $\mathcal{R}$ and transmission $\mathcal{T}$ coefficients are analyzed for the case $\varphi_0=-\pi/4$ and $k_y=0$. The absolute values of $\mathcal{R}$ and $\mathcal{T}$ as functions of $k_x$ are shown on panels (a) and  (b), respectively. The solid and dashed lines correspond to the RH- and LH-polarized magnons, respectively. Panels (c) and (d) illustrate the evolution of the phase shift $\delta(k_x)$ for the RH- and LH-polarized magnons, respectively.}
    \label{fig:RT}
\end{figure}
For the sake of simplicity, we limit ourselves to the case of normal incidence of a spin wave on a DW, i.e., $k_y=0$. In this case, if $\varepsilon\ne0$, in the limit $|k_x|\ll|\varepsilon|\ll1$ we obtain $\mathcal{R}\approx -(1\pm2ik_x\varepsilon^{-1})$, and $\mathcal{T}\approx\pm2ik_x\varepsilon^{-1}$, where signs `$+$' and `$-$' correspond to RH- and LH-polarized magnons, respectively. In the opposite limit of large $k_x$, the estimation \eqref{eq:R-approx} should be used. Remarkably, the reflection and transmission coefficients depend on the magnon polarization, RH or LH. This polarization asymmetry is also demonstrated in Fig.~\ref{fig:RT}(a,b). Since $\mathcal{R}(-k_x)=\mathcal{R}^*(k_x)$, and $\mathcal{T}(-k_x)=\mathcal{T}^*(k_x)$~\footnote{This is correct for $k_y=0$ only.}, without loss of generality we consider only $k_x>0$ in Fig.~\ref{fig:RT}(a,b). Note that a rotation of the DW by $\pi/2$, as well as the flip of the sign of $\varepsilon$, interchanges the results for the RH and LH polarizations. In what follows, we fix the sign $\varepsilon>0$ without loss of generality. Since the transmission coefficient $\mathcal{T}$ (as well as $\mathcal{R}$) does not depend on the DW topological charge $p$ (in the absence of the magnetic field), one can use an array of A$\ell$M DWs as a polarizing filter. Note that DWs in an array have alternating topological charges. 

Due to the nonzero reflection coefficient, one expects much more effective propulsion of A$\ell$M DW by magnons, as compared to the AFM DW~\cite{Kim14a,Tveten14}.

Another exotic feature of the magnon scattering on an A$\ell$M DW is illustrated in Figs.~\ref{fig:RT}(c,d). They show the phase shift $\delta(k_x)=\text{arg}(\mathcal{T}(k_x))$, which the transmitted wave gains relative to the incident wave. Using \eqref{eq:TR}, one obtains the following asymptotic behavior $\delta(k_x)\sim u/k_x$ in the limit $k_x\to\infty$. In this limit, the amplitude of the scattering potential $u$ differs drastically for different polarizations, namely $u\simeq1/2$ and $u\simeq3\varepsilon^2k_x^2/[2(1-\varepsilon^2)]$ for RH- and LH-polarized spin waves, respectively. As a result, for the RH polarization, one has $\delta(\infty)=0$ and the scattering process fulfills the one-dimensional Levinson's theorem~\footnote{Levinson's theorem for 3D and 1D theories differs in the treatment of half-bound states. Namely, the $1/2$ term is present when there is no half-bound state, unlike in 3D~\cite{Kiers96}.}, i.e., $\delta(0)-\delta(\infty)=\pi$ and $\delta(0)-\delta(\infty)=\pi/2$ for the presence and absence of the half-bound state, respectively. Note that for $\varepsilon\ge0$, a single RH bound state is present and a half-bound state appears for $\varepsilon=0$ only. For the LH polarization, $\delta(\infty)\to\infty$, and Levinson's theorem is violated. The reason for the theorem violation is the energy dependence of the scattering potential, namely, its infinite growth for $k_x\to\infty$.


Finally, we note that in general case, the angle of reflection $\theta_{\text{refl}}=\arctan[k^{\text{refl}}_y/(-k^{\text{refl}}_x)]$ is not equal to the angle of incidence $\theta_{\text{inc}}=\arctan[k^{\text{inc}}_y/k^{\text{inc}}_x]$, where $k^{\text{refl}}_\alpha$ and $k^{\text{inc}}_\alpha$ denotes the vave-vector components of the reflected and incident waves, respectively. This happens due to the shift of $x$-component of the wave vector $k^{\text{refl}}_x=- (k^{\text{inc}}_x-2q)$ in the reflected wave. The latter is a direct consequence of the anisotropic magnon dispersion relation and the requirement $k^{\text{refl}}_y=k^{\text{inc}}_y=k_y$, which follows from the translational invariance in the $y$-direction, i.e, along the DW. In other words, the shift of $k^{\text{refl}}_x$ is such that the relation $\omega(k^{\text{inc}}_x,k_y)=\omega(k^{\text{refl}}_x,k_y)$ holds. The effect strongly depends on the DW orientation. It is maximal for a nodal DW orientation. In this case, the angles of incidence and reflection are related by the following expression $\tan\theta_{\text{refl}}=\tan\theta_{\text{inc}}/[1\pm2\varepsilon\omega\tan\theta_{\text{inc}}]$, where signs `$+$' and `$-$' in denominator correspond to the cases $\varphi_0=0$ and $\varphi_0=\pi/2$, respectively. While for the diagonal DW orientation $\varphi_0=\pm\pi/4$, one has $\theta_{\text{refl}}=\theta_{\text{inc}}$.





\section{Conclusions}
Based on the Lagrangian of the nonlinear sigma model recently developed for $d$-wave altermagnets~\cite{Gomonay24a,Yershov25,Vakili25}, we analyze the propagation of bound states along a static DW. In contrast to an AFM DW, which possesses a single bound state (translational mode), an A$\ell$T DW can have additional bound states. The number of the new A$\ell$T-induced resonances depends on the DW orientation relative to the crystallographic directions and on the strength of the A$\ell$T parameter $\varepsilon$. Even for an infinitely small value of $\varepsilon$, an additional bound state appears if the DW is oriented not in a nodal direction. The latter is important for practical realizations, since $\varepsilon\ll1$ for presently known A$\ell$Ts. The A$\ell$T-induced eigenstates exhibit the exotic property of focusing on the DW as their wave vector increases. The latter effect can be useful for miniaturizing magnonic devices that utilize DWs as waveguides.

Next, we analyze the influence of the magnetic field applied along the easy axis for the case of a nodal DW orientation. We find that the field results in a strong asymmetry in the spatial distribution of the eigenstate intensity relative to the DW. In addition to this, the field limits the range of possible wave vectors for the bound state. For an arbitrary orientation of the A$\ell$T DW, the magnetic field significantly changes the DW profile compared to the standard AFM solution. This particular aspect deserves further study.

Finally, we consider properties of the scattering states. We find that the reflection coefficient is not zero if the A$\ell$M DW is oriented not in a nodal direction. This contrasts with static AFM DWs, which are known to create a reflectionless scattering potential for spin waves. Thus, one should expect much more effective propulsion of A$\ell$M DW by magnons, as compared to the AFM DW~\cite{Kim14a,Tveten14}. We also indicate the violation of Levinson's theorem, resulting from the energy dependence of the scattering potential.


The best platform for an experimental observation of the new localized states is an insulating $d$-wave altermagnet in which the magnon-electron damping is suppressed, and the linewidth is small. Possible candidates are thin films of $\kappa$-Cl~\cite{Eto25} or La$_2$O$_3$Mn$_2$Se$_2$~\cite{Garcia-Gassull26} deposited on a heavy-metal substrate such that the additional easy-axial anisotropy is induced on the interface.

\section{Acknowledgments}
This work was supported by the German Federal Ministry of Research, Technology and Space (BMFTR) through the GU-QuMat project (01DK24008). V.K.~acknowledges financial support from the Deutsche
Forschungsgemeinschaft (DFG, German Research Foundation) through the W{\"u}rzburg-Dresden Cluster of Excellence ctd.qmat – Complexity, Topology and Dynamics
in Quantum Matter (EXC 2147, project-id 390858490).
\appendix

\section{Dispersion relation for the uniform ground state}\label{app:GS-DW}
In order to find linear excitation above a static solution $\vec{n}_0=\mathrm{p}\vec{e_z}$ with $\mathrm{p}=\pm1$ in magnetic field $\vec{b}=b\vec{e}_z$, we first derive the spin-wave Lagrangian $\mathscr{L}_{\text{sw}}$. To this end, we substitute \eqref{eq:HP} into Lagrangian \eqref{eq:L-magnetic}, and account only for the harmonic part 
\begin{equation} \label{eq:L-SW-dim}
\begin{split}
   \mathscr{L}_{\text{sw}}= \frac12\bigl[&|\dot\psi|^2-|\vec{\nabla}\psi|^2-(1-b^2)|\psi|^2+2i\mathrm{p}b\,\psi\dot{\psi}^*\\
   &+\varepsilon(i\dot{\psi}-\mathrm{p}b\psi)\hat{\mathfrak{D}}_{\{x,y\}}\psi^*+\text{c.c.}\bigr]
   \end{split}
\end{equation}
The equation of motion generated by Lagrangian \eqref{eq:L-SW-dim} is
\begin{equation}\label{eq:motion-SW}
    \ddot{\psi}-\nabla^2\psi+(1-b^2)\psi+2i\mathrm{p}b\dot{\psi}-i\varepsilon\hat{\mathfrak{D}}_{\{x,y\}}\dot\psi+\varepsilon\mathrm{p}b\hat{\mathfrak{D}}_{\{x,y\}}\psi=0.
\end{equation}
It describes a planar wave $\psi = \Psi e^{i(k_xx+k_yy-\omega t)}$  propagating on the top of the saturated state with the dispersion relation
\begin{equation}\label{eq:disp}
    \omega(\vec{k})= \mathrm{p} b \pm\sqrt{\omega_{\text{afm}}^2(\vec{k}) +\frac{\varepsilon^2\omega^2_{\mathrm{alt}}(\vec{k})}{4}}+\frac{\varepsilon\omega_{\mathrm{alt}}(\vec{k})}{2},
\end{equation}
where $\omega_{\text{afm}}=\sqrt{1+k_x^2+k_y^2}$ and $\omega_{\text{alt}}=2\cos(2\varphi_0)k_xk_y-\sin(2\varphi_0)(k_x^2-k_y^2)$. In the dimension units, dispersion relation \eqref{eq:disp-sw} obtains form \eqref{eq:disp-sw}.

\section{Bound states of a DW without magnetic field}\label{sec:hypergeom-theory}

The equation of motion generated by Lagrangian~\eqref{eq:L-DW-0} is
\begin{equation}
\begin{split}
    -\ddot{\psi}+\vec{\nabla}^2\psi&-\psi+i\varepsilon\hat{\mathfrak{D}}_{\{x,y\}}\dot\psi\\
    &+\frac{2}{\cosh^2x}\left(\psi-\frac{i\varepsilon}{4}\sin2\varphi_0\dot{\psi}\right)=0.
\end{split}    
\end{equation}
The substitution $\psi=\Psi(x)e^{iqx}e^{i(k_yy-\omega t)}$ results in Eq.~\eqref{eq:Shrod-0} for the amplitude $\Psi(x)$ if one chooses $q =-\varepsilon\omega k_y\cos2\varphi_0/ (1-\varepsilon\omega\sin2\varphi_0)$ . Equation \eqref{eq:Shrod-0}, as any homogeneous linear differential equation of the second order with at most three distinct singularities, all regular,  can be transformed into a hypergeometric equation~\cite{NIST10}. To this end, we first make the substitution $\Psi(x)=f(x)\sech^\kappa x$, which leads to equation
\begin{equation}\label{eq:f}
    f''_{xx}-2\kappa\tanh{x}f'_x+\frac{u-\kappa-\kappa^2}{\cosh^2x}f=0.
\end{equation}
The subsequent change of variables 
\begin{equation}\label{eq:z}
 z =\frac{1-\tanh{x}}{2}   
\end{equation}
transforms~\eqref{eq:f} to the hypergeometric differential equation~\cite{NIST10}
\begin{equation}\label{eq:hg}
   z(1-z)f''_{zz}+\left(c-(a+b+1)z\right)f'_{z}-abf=0 
\end{equation}
with coefficients $a, \ b,$ and $c$ defined as 
\begin{equation}\label{eq:coef-0}
    a=\kappa+\frac12+\sqrt{u+\frac14},\ b=\kappa+\frac12-\sqrt{u+\frac14},\ c=1+\kappa.
\end{equation}
Thus, the solution of Eq.~\eqref{eq:Shrod-0} is
\begin{equation}\label{eq:Psi-sol}
    \Psi(x)=\mathcal{C}\frac{_2F_1\left(a,b;c;z(x)\right)}{\cosh^\kappa x},
\end{equation}
where $\mathcal{C}$ is an arbitrary constant and $0<z(x)<1$ is defined above. In the following, we consider only bounded solutions with $\kappa\ge0$.

Let us consider the asymptotics of the solution \eqref{eq:Psi-sol} in the limits $x\to\pm\infty$. To this end, we use that 
\begin{equation}\label{eq:F21-lim0}
 _2F_1(a,b;c;\xi)=1+\mathcal{O}(\xi),\quad|\xi|\ll1.   
\end{equation}
In the limit $x\to\infty$, one has $z\to0$, and with \eqref{eq:F21-lim0} we obtain $\Psi\sim\mathcal{C}2^\kappa e^{-\kappa x}$. In the opposite limit $x\to-\infty$, we have $z\to1$. In order to utilize \eqref{eq:F21-lim0} we use relation~\cite{Bateman53}
\begin{equation}\label{eq:Kummer}
    \begin{split}
    &_2F_1(a,b;c;z)=\\
    &=\frac{\Gamma(c)\Gamma(c-a-b)}{\Gamma(c-a)\Gamma(c-b)}{}_2F_1(a,b;a+b-c+1;1-z)\\
    &+(1-z)^{c-a-b}\frac{\Gamma(c)\Gamma(a+b-c)}{\Gamma(a)\Gamma(b)}\\
    &\times{}_2F_1(c-a,c-b;c-a-b+1;1-z).
    \end{split}
\end{equation}
Taking into account that $c-a-b=-\kappa$, in the limit $x\to-\infty$, one obtains $(1-z)^{c-a-b}\sim e^{-2\kappa x}$. Now, with the use of \eqref{eq:Kummer} and \eqref{eq:F21-lim0} we derive asymptotics
\begin{equation}
    \Psi\sim\mathcal{C}2^{\kappa}\left[\frac{\Gamma(c)\Gamma(c-a-b)}{\Gamma(c-a)\Gamma(c-b)}e^{\kappa x}+\frac{\Gamma(c)\Gamma(a+b-c)}{\Gamma(a)\Gamma(b)}e^{-\kappa x}\right].
\end{equation}
The condition for the existence of a bound state is the vanishing of the coefficient at $e^{-\kappa x}$ (note that $x<0$). This is possible if a gamma function in the denominator has a pole. Since $a$ is positive by definition, the function $\Gamma(a)$ does not have poles. On the other hand, $\Gamma(b)$ has poles under the condition $b=-n$, where $n=0,1,2,\dots$ The latter condition is equivalent to \eqref{eq:bound-condition}.

\section{Bound states of nodal DW in the presence of the magnetic field}\label{sec:hypergeometric-magnetic}

Lagrangian \eqref{eq:L-h} produces the equation of motion 
\begin{equation}\label{eq:motion-magnetic}
    \begin{split}
   &- \ddot{\psi} + \nabla^2 \psi -  \varkappa^2 (1 - 2 \sech^2{\varkappa x}) \psi   + 2 i \varepsilon \dot{\psi}_{xy} \\
   & + 2 p b \tanh{\varkappa  x} (i\dot{\psi} + \varepsilon \psi_{xy}) +p \varepsilon b \varkappa \sech^2{\varkappa x} \  \psi_y = 0.
\end{split}
\end{equation}

Next, we utilize a similar change of variables as in Appendix ~\ref{sec:hypergeom-theory}, namely $z = \frac{1}{2} (1-\tanh{\varkappa x})$.
This allows us rewrite Eq.~\eqref{eq:phi0simp} as
\begin{equation}\label{eq:Fuchsian}
    z(1-z)\partial^2_{zz}\Psi + (1-2z)\partial_z\Psi + \left[\beta- \frac{\alpha-\gamma}{4 z} - \frac{\alpha+\gamma}{4(1-z)}   \right] \Psi = 0,
\end{equation}
Now, by means of the substitution $\Psi(z) = z^{\kappa_+/2}(1-z)^{\kappa_-/2} f(z)$ with $\kappa_\pm=\sqrt{\alpha\pm\gamma}$, we reduce Eq.~\eqref{eq:Fuchsian} to the standard hypergeometric differential equation~\eqref{eq:hg} with $a=\frac12[\kappa_++\kappa_-+1]+(\beta+\frac14)^{1/2}$, $b=\frac12[\kappa_++\kappa_-+1]-(\beta+\frac14)^{1/2}$, and $c=1+\kappa_+$. Thus, the solution of Eq.~\eqref{eq:Fuchsian} is $\Psi=\mathcal{C}z^{\kappa_+/2}(1-z)^{\kappa_-/2}{}_2F_1(a,b;c;z)$.

Let us now investigate the asymptotics of the solution at infinitely large distances from the DW. In the limit $x\to\infty$, one obtains $z\sim e^{-2\varkappa x}$. And with the use of \eqref{eq:F21-lim0}, we estimate $\Psi\sim \mathcal{C}e^{-\kappa_+\varkappa x}$. In the opposite limit $x\to-\infty$, we have $1-z\sim e^{2\varkappa x}$ and $z\sim1$. Now, with the use of \eqref{eq:Kummer} and \eqref{eq:F21-lim0}, we obtain
\begin{equation}
    \Psi\!\sim\!\mathcal{C}\left[\frac{\Gamma(c)\Gamma(c-a-b)}{\Gamma(c-a)\Gamma(c-b)}e^{\kappa_-\varkappa x}\!+\frac{\Gamma(c)\Gamma(a+b-c)}{\Gamma(a)\Gamma(b)}e^{-\kappa_-\varkappa x}\!\right].
\end{equation}
The condition for the existence of a bound solution is the vanishing of the coefficient at $e^{-\kappa_-\varkappa x}$ (note that $x<0$), i.e., $b=-n$, where $n=0,1,\dots$

\section{Scattering states}\label{app:sct}

We look for solutions of Eq.~\eqref{eq:Shrod-0} that have asymptotics at $x\to\pm\infty$ given by planar waves. To this end, we look for solutions in the form $\Psi=(\cosh x)^{i\tilde{k}_x}f(x)$, where $\tilde{k}_x$ is a real number~\footnote{Here we follow the procedure described in Ref.~\onlinecite{Lamb80}}. Under the condition
\begin{equation}\label{eq:cnd-disp}
    \tilde{k}_x^2+\kappa^2=0,
\end{equation}
Eq.~\eqref{eq:Shrod-0} obtains form of Eq.~\eqref{eq:f}, in which $\kappa=-i\tilde{k}_x$. And in terms of the $z$-coordinate defined in~\eqref{eq:z}, Eq.~\eqref{eq:f} is reduced to the standard hypergometric equation \eqref{eq:hg}, in which the coefficients $a$, $b$, $c$ are the same as in~\eqref{eq:coef-0} with $\kappa=-i\tilde{k}_x$.
In the limit $x\to\infty$, we obtain the asymptotic behavior $\Psi\sim 2^{-i\tilde{k}_x}e^{i\tilde{k}_xx}$, and $\psi\sim 2^{-i\tilde{k}_x}e^{i(q+\tilde{k}_x)x}$. By introducing $k_x=q+\tilde{k}_x$, we obtain solution~\eqref{eq:psi-sct}. Important that the condition~\eqref{eq:cnd-disp} written in the form $(k_x-q)^2+\kappa^2=0$, coincides with the dispersion relation~\eqref{eq:disp} for $b=0$. I.e., the dispersion relation of the scattering states is the same as for the magnons excited on the top of a uniform ground state.

Using relation~\eqref{eq:Kummer} and the property~\eqref{eq:F21-lim0} of the hypergeometric function, we find the asymptotics at $x\to-\infty$ and derive the transmission and reflection coefficients~\eqref {eq:TR}.

\end{document}